\documentclass{elsart}

 \usepackage{graphicx}
\usepackage{amssymb}

\journal{Physics Letters A}

\newcommand{\id}{{\sf 1 \hspace{-0.3ex}
\rule{0.1ex}{1.52ex}\rule[-.01ex]{0.3ex}{0.1ex}}}

\newcommand{\shalf}{{\textstyle\frac{1}{\sqrt{2}}}}
\newcommand{\ket}[1]{|#1\rangle}
\newcommand{\bra}[1]{\langle#1|}
\newcommand{\braket}[2]{\langle#1|#2\rangle}
\newcommand{\ketbra}[2]{|#1\rangle\langle#2|}
\newcommand{\proj}[1]{\ket{#1}\bra{#1}}

\begin{document}

\begin{frontmatter}



 \title{Cluster-type entangled coherent states}


 \author[unicamp]{P. P. Munhoz},
 \author[uepg,unicamp]{F. L. Semi\~ao\corauthref{cor1}},
 \ead{semiao@uepg.br}
 \author[unicamp]{A. Vidiella-Barranco}, \and
 \author[unicamp]{J. A. Roversi}
 \corauth[cor1]{Corresponding author at: Departamento de F\'isica, Universidade Estadual de Ponta Grossa - Campus Uvaranas, 84030-900 Ponta Grossa, PR, Brasil, Phone: +55 (42)3220-3047, Fax: +55 (42)3220-3042.}
 \address[unicamp]{Instituto de F\'\i sica ``Gleb Wataghin'' - Universidade Estadual de Campinas, Unicamp - 13083-970 Campinas, SP, Brasil}
 \address[uepg]{Departamento de F\'isica, Universidade Estadual de Ponta Grossa - Campus Uvaranas, 84030-900 Ponta Grossa, PR, Brasil}

\begin{abstract}
% Text of abstract
We present the cluster-type entangled coherent states (CTECS) and discuss their properties. A cavity QED generation scheme using suitable choices of atom-cavity interactions, obtained via detunings adjustments and the application of classical external fields, is also presented. After the realization of simple atomic measurements, CTECS representing nonlocal electromagnetic fields in separate cavities can be generated.
\end{abstract}

\end{frontmatter}

% main text
%\section{}
%\label{}

Multipartite entangled states are vital for the full exploration of quantum computation and communication protocols. A special kind of multipartite entangled states called cluster states \cite{persistent} have attracted much attention due to their potential applications. For instance, such states form the basis of the one-way quantum computing model \cite{one-way1,one-way2,one-way3}, an alternative to the conventional circuit approach \cite{NielsenChuang}. Experimental generation of cluster states encoded in the polarization state of four photons and their application in the implementation of Glover's search algorithm have already been reported \cite{WaltherC}. More fundamental issues as conceptual foundations of quantum mechanics have been studied in the scope of cluster states as well \cite{Scarani}. Quantum information protocols including teleportation and dense coding have been proposed recently \cite{tdc}. The physical implementation of such states is then of major importance and several different physical systems have been considered. We mention proposals involving photons \cite{photons1,photons2,photons3,photons4,photons5}, trapped ions \cite{ZhengC1,ZhengC2,ZhengC3}, cavity QED  \cite{cavityQED1,cavityQED2,cavityQED3,cavityQED4,cavityQED4b,cavityQED5,cavityQED6,cavityQED7,cavityQED8,cavityQED9}, hybrid cavity-QED/linear optics setups \cite{cqedlin1,cqedlin2}, and superconducting quantum circuits \cite{supercondC1,supercondC2,supercondC3}, just to name a few.

Cluster states have been conceived in finite discrete systems, typically in a tensor product structure of the type $\mathbb{C}^2\otimes\ldots\otimes\mathbb{C}^2$ (qubits). Motivated by the notable success achieved by the use of infinite-dimensional continuous variable systems in quantum information protocols \cite{cvprotocols1,cvprotocols2,cvprotocols3,cvprotocols4}, we put forward an extension of the usual qubit-based cluster state to the continuous case. In order to do that, we propose the use of coherent states \cite{coherentst} since their generation and manipulation is well established in various experimental setups such as microwave cavities \cite{harochecoh} and trapped ions \cite{winelandcoh}. Additionally, coherent states have been previously considered for quantum teleportation \cite{teleportcoh1,teleportcoh2}, quantum information processing \cite{gatescoh1,gatescoh2}, and tests of local realism \cite{localrealcoh1,localrealcoh2}. All those applications and interesting features are determinant for the choice of coherent states for our proposal, the CTECS. We mention that other types of continuous-variable cluster states based on different states (not coherent) have been already discussed in the literature \cite{cvcluster1,cvcluster2,cvcluster3,cvcluster4}.

The basic states used in this paper are the coherent states $\ket\alpha$ and $\ket{{-}\alpha}$, defined as $\ket{{\pm}\alpha}=\hat{D}(\pm\alpha)\ket{0}$ (displaced vacuum states), where $\hat{D}(\pm\alpha)=\e^{\pm(\alpha\hat{a}^\dag-\alpha^\ast\hat{a})}$. They have the useful property that the overlap $\braket{\alpha}{{-}\alpha}$ decays exponentially with $\alpha$. For $\alpha=3$, for instance, the overlap is about $10^{-8}$, and for practical applications one may use them for encoding a qubit $\ket\phi$ in $\mathbb{C}^2$ as \cite{gatescoh1,gatescoh2}
\begin{equation}
\ket\phi=\cos(\theta/2)\ket\alpha+e^{i\psi}\sin(\theta/2)\ket{{-}\alpha}. \label{qubit}
\end{equation}
These states $\ket\alpha$ and $\ket{{-}\alpha}$ can be discriminated (with high probability) by a simple measurement scheme involving a $50-50$ beam splitter as explained in \cite{gatescoh1,gatescoh2}. Also, starting from a coherent state $\ket\alpha$, an arbitrary qubit state (\ref{qubit}) may be prepared, up to a global phase, using phase-shifters, beam-splitters, nonlinear medium, and auxiliary coherent state modes \cite{gatescoh1,gatescoh2}. In other words, arbitrary one-qubit gates may be obtained combining optical components and fields. For a universal set of quantum gates to be complete, a controlled two-qubit gate is necessary besides arbitrary manipulations of one qubit. This can be achieved by employing a quantum teleportation protocol \cite{gatescoh1,gatescoh2,teleportgatescoh}. From these facts, we see that a qubit involving coherent states is a potential alternative choice for information encoding and manipulation in quantum computing.

One can then take a step ahead and look into multipartite systems described by entangled coherent states. The simplest case is the bipartite scenario. The attempt is again trying to use infinite-dimensional systems (modes) to describe a state in a finite-dimensional space, in this case $\mathbb{C}^2\otimes\mathbb{C}^2$. Previous papers \cite{teleportcoh1,teleportcoh2,localrealcoh1,localrealcoh2,ECS} have proposed and investigated the properties of the Bell-type entangled coherent states of the form
\begin{eqnarray}\label{quasiBell}
\ket{\Phi^\pm_\alpha}&=&N_\pm(\ket{\alpha,\alpha}\pm\ket{{-}\alpha,{-}\alpha}),\nonumber\\
\ket{\Psi^\pm_\alpha}&=&N_\pm(\ket{\alpha,{-}\alpha}\pm\ket{{-}\alpha,\alpha}),
\end{eqnarray}
where $N_\pm=[2(1\pm\e^{-4|\alpha|^2})]^{-1/2}$. They are named quasiBell states in analogy with the Bell states $\ket{\Phi^\pm}=\shalf(\ket{00}\pm\ket{11})$ and $\ket{\Psi^\pm}=\shalf(\ket{01}\pm\ket{10})$ which form an orthonormal basis in $\mathbb{C}^2\otimes\mathbb{C}^2$ denoted here as $\mathcal{B}_{\rm Bell}$. The states (\ref{quasiBell}) cannot be formally called Bell states since they are not mutually orthogonal for finite values of $\alpha$. However, the overlap $\braket{\alpha}{{-}\alpha}$ tends to zero very rapidly with the increasing of $\alpha$, as mentioned before. The states (\ref{quasiBell}) constitute a non-orthonormal basis denoted here as $\mathcal{B}^\alpha_{\rm Bell}$.

We are interested in generalizations involving coherent states of important qubit-based multipartite states. In the tripartite case the GHZ and W states based on coherent states have been shown to violate Mermin's version of the Bell inequality \cite{jeongmermin}. Now we would like to address $4$-partite entangled states in the form of the usual cluster state \cite{persistent}
\begin{eqnarray}\label{ucl}
\ket{{\rm CLUSTER}^+}=\half(\ket{0000}+\ket{0011}+\ket{1100}-\ket{1111}).
\end{eqnarray}
In analogy with the Bell states in the two-qubit case, we \emph{define} now an orthonormal basis $\mathcal{B}_{\rm C}$ for the $4$-qubit setting. This basis contains the elements
\begin{eqnarray}\label{clusterbasis}
&&\ket{{\rm CLUSTER}^\pm}=\half(\pm\ket{0000}+\ket{0011}+\ket{1100}\mp\ket{1111}),\nonumber\\
&&\ket{{\rm C}^\pm}=\half(\ket{0000}\pm\ket{0011}\mp\ket{1100}+\ket{1111}),\nonumber\\
&&\ket{{\rm L}^\pm}=\half(\pm\ket{0001}\mp\ket{0010}+\ket{1101}+\ket{1110}),\nonumber\\
&&\ket{{\rm U}^\pm}=\half(\ket{0001}+\ket{0010}\pm\ket{1101}\mp\ket{1110}),\nonumber\\
&&\ket{{\rm S}^\pm}=\half(\pm\ket{0100}+\ket{0111}\mp\ket{1000}+\ket{1011}),\nonumber\\
&&\ket{{\rm T}^\pm}=\half(\ket{0100}\pm\ket{0111}+\ket{1000}\mp\ket{1011}),\nonumber\\
&&\ket{{\rm E}^\pm}=\half(\pm\ket{0101}+\ket{0110}+\ket{1001}\mp\ket{1010}),\nonumber\\
&&\ket{{\rm R}^\pm}=\half(\ket{0101}\pm\ket{0110}\mp\ket{1001}+\ket{1010}).
\end{eqnarray}
Note that the elements of $\mathcal{B}_{\rm C}$ defined in (\ref{clusterbasis}) are locally equivalent via the application of bit-flips. Consequently, they possess the same amount of entanglement. The reader may easily show that by tracing out $3$ particles, the remaining one is left in a maximally mixed state $\hat\rho=\id/d$, where $d=2$ for qubits. The map between elements in $\mathcal{B}_{\rm C}$ just makes use of bit-flips and this is a remarkable feature absent in $\mathcal{B}_{\rm Bell}$. Actually, not even a tripartite orthonormal basis defined using the GHZ states \cite{Rigolin} has such a property, i.e., more than one type of local operation must be used. We now \emph{propose} a generalization of such cluster states by using the coherent states encoding. We call them cluster-type entangled coherent states (CTECS). The state (\ref{ucl}) becomes
\begin{eqnarray}\label{quasiCluster}
\ket{{\rm CLUSTER}^+_\alpha}&=&\half(\ket{\alpha,\alpha,\alpha,\alpha}+\ket{\alpha,\alpha,{-}\alpha,{-}\alpha}+\ket{{-}\alpha,{-}\alpha,\alpha,\alpha}\\\nonumber
&&-\ket{{-}\alpha,{-}\alpha,{-}\alpha,{-}\alpha}),
\end{eqnarray}
and, in complete analogy with the quasi-Bell basis, we use $\mathcal{B}_{\rm C}$ to \emph{define} the coherent state generalized basis (CTECS basis) $\mathcal{B}^\alpha_{\rm C}$ composed of the elements
\begin{eqnarray}\label{CTECSbasis}
&&\ket{{\rm CLUSTER}^\pm_\alpha}=\half(\pm\ket{\alpha,\alpha,\alpha,\alpha}+\ket{\alpha,\alpha,{-}\alpha,{-}\alpha}+\ket{{-}\alpha,{-}\alpha,\alpha,\alpha}\mp\ket{{-}\alpha,{-}\alpha,{-}\alpha,{-}\alpha}),\nonumber\\
&&\ket{{\rm C}^\pm_\alpha}=\half(\ket{\alpha,\alpha,\alpha,\alpha}\pm\ket{\alpha,\alpha,{-}\alpha,{-}\alpha}\mp\ket{{-}\alpha,{-}\alpha,\alpha,\alpha}+\ket{{-}\alpha,{-}\alpha,{-}\alpha,{-}\alpha}),\nonumber\\
&&\ket{{\rm L}^\pm_\alpha}=\half(\pm\ket{\alpha,\alpha,\alpha,{-}\alpha}\mp\ket{\alpha,\alpha,{-}\alpha,\alpha}+\ket{{-}\alpha,{-}\alpha,\alpha,{-}\alpha}+\ket{{-}\alpha,{-}\alpha,{-}\alpha,\alpha}),\nonumber\\
&&\ket{{\rm U}^\pm_\alpha}=\half(\ket{\alpha,\alpha,\alpha,{-}\alpha}+\ket{\alpha,\alpha,{-}\alpha,\alpha}\pm\ket{{-}\alpha,{-}\alpha,\alpha,{-}\alpha}\mp\ket{{-}\alpha,{-}\alpha,{-}\alpha,\alpha}),\nonumber\\
&&\ket{{\rm S}^\pm_\alpha}=\half(\pm\ket{\alpha,{-}\alpha,\alpha,\alpha}+\ket{\alpha,{-}\alpha,{-}\alpha,{-}\alpha}\mp\ket{{-}\alpha,\alpha,\alpha,\alpha}+\ket{{-}\alpha,\alpha,{-}\alpha,{-}\alpha}),\nonumber\\
&&\ket{{\rm T}^\pm_\alpha}=\half(\ket{\alpha,{-}\alpha,\alpha,\alpha}\pm\ket{\alpha,{-}\alpha,{-}\alpha,{-}\alpha}+\ket{{-}\alpha,\alpha,\alpha,\alpha}\mp\ket{{-}\alpha,\alpha,{-}\alpha,{-}\alpha}),\nonumber\\
&&\ket{{\rm E}^\pm_\alpha}=\half(\pm\ket{\alpha,{-}\alpha,\alpha,{-}\alpha}+\ket{\alpha,{-}\alpha,{-}\alpha,\alpha}+\ket{{-}\alpha,\alpha,\alpha,{-}\alpha}\mp\ket{{-}\alpha,\alpha,{-}\alpha,\alpha}),\nonumber\\
&&\ket{{\rm R}^\pm_\alpha}=\half(\ket{\alpha,{-}\alpha,\alpha,{-}\alpha}\pm\ket{\alpha,{-}\alpha,{-}\alpha,\alpha}\mp\ket{{-}\alpha,\alpha,\alpha,{-}\alpha}+\ket{{-}\alpha,\alpha,{-}\alpha,\alpha}).
\end{eqnarray}

We now discuss some simple properties of $\mathcal{B}^\alpha_{\rm C}$. First, it is worth noticing that the elements in the basis can still be interconverted via generalized  bit-flip operations now defined using the coherent states encoding as \cite{powercat}
\begin{eqnarray}\label{bitflip}
&&\hat{X}\ket\alpha=\hat{P}(\pi)\ket\alpha=\ket{{-}\alpha},\nonumber\\
&&\hat{X}\ket{{-}\alpha}=\hat{P}(\pi)\ket{{-}\alpha}=\ket\alpha,
\end{eqnarray}
where $\hat{P}(\pi)=\e^{i\pi{}\hat{n}}$ is the parity operator, with $\hat{n}$ being the number operator of a bosonic field. Thus, it follows a second property which states that all elements in $\mathcal{B}^\alpha_{\rm C}$ have the same $\alpha$-dependent entanglement content. The reduced state of one subsystem (obtained by tracing out the other three states) is now a function of $\alpha$
\begin{equation}\label{rhoalpha}
\hat\rho_\alpha=\half[\proj{\alpha}(1+\e^{-4|\alpha|^2})+\proj{{-}\alpha}(1-\e^{-4|\alpha|^2})].
\end{equation}
From equation (\ref{rhoalpha}), one can see that in the limit of $\alpha\rightarrow\infty$, the state $\hat\rho_\alpha$ tends to a maximally mixed state in $\mathbb{C}^2$ because the overlap $\braket{\alpha}{{-}\alpha}$ approaches zero. This means that (\ref{clusterbasis}) may be formally obtained from (\ref{CTECSbasis}) in that limit.

\begin{figure}[h]
\centering\includegraphics[width=0.8\columnwidth]{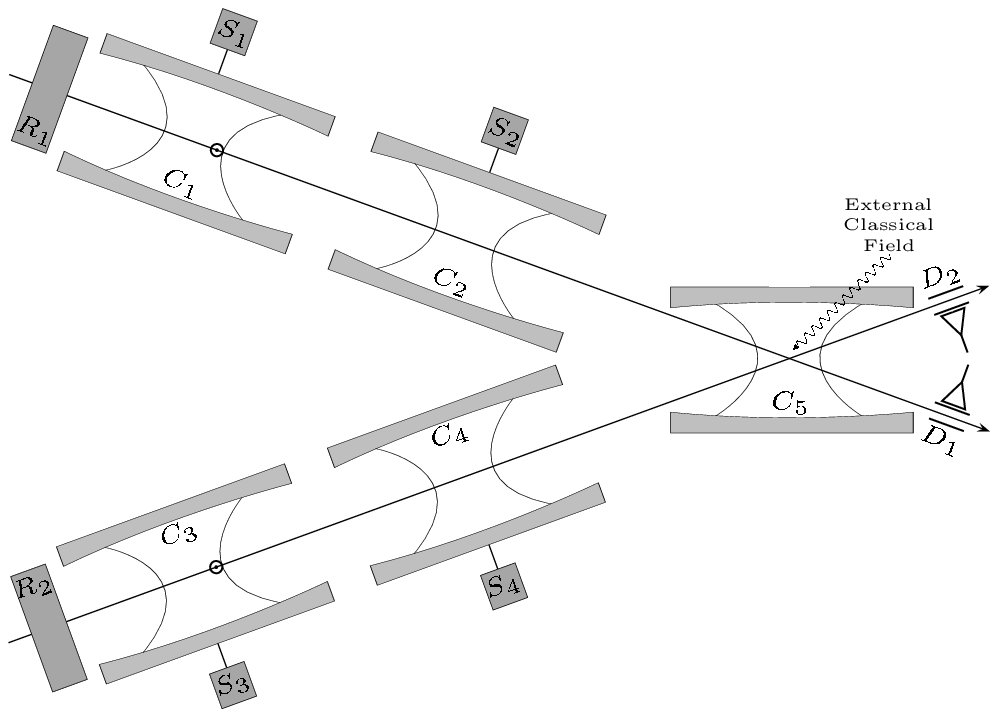}
\caption{Sketch of the proposed setup. Each flying atom crosses a Ramsey zone and two cavities which sustain coherent fields of the same amplitude. Both atoms enter a fifth cavity whose role is to apply a quantum phase gate between the atoms before measurement in $D_1$ and $D_2$.\label{setup}}
\end{figure}

We now present a cavity-QED implementation of the CTECS. The scheme we have in mind is depicted in Fig.\ref{setup}. Consider two identical two-level atoms prepared in the pure state $\ket\psi_{\rm a}=\ket{g}_1\ket{g}_2=\ket{gg}$, with $\ket{e}$ and $\ket{g}$ being the usual electronic excited and ground states, respectively. As we are going to describe below, the atoms are sent to cross high-Q cavities sustaining coherent states such that the initial state of the whole system is $\ket\psi_{\rm af}=\ket\psi_{\rm a}\ket\alpha_1\ket\alpha_2\ket\alpha_3\ket\alpha_4=\ket{gg}\ket{\alpha,\alpha,\alpha,\alpha}$. As depicted in Fig.\ref{setup}, the atom $1$ crosses cavities $C_1$ and $C_2$ in sequence and atom 2 crosses cavities $C_3$ and $C_4$.

First, each atom is sent to separate Ramsey zones $R_1$ and $R_2$ where their electronic state is rotated by a $\pi/2$-pulse which prepares each atom in the coherent superposition $\shalf(\ket{g}+\ket{e})$. The atoms then interact dispersively with the fields in cavities $1$, $2$, $3$ and $4$ (see Fig. \ref{setup}). In the dispersive regime, there is no direct energy exchange between field and atom and this regime is achieved when the detuning between frequencies of atom and cavity field $\Delta$ is much larger than their coupling constant $g$. However, the atom and field are still coupled and atomic and field phases change in time. The effective dispersive interaction Hamiltonian (RWA) can be easily derived \cite{DispersiveJCM1,DispersiveJCM2} and it is given by
\begin{equation}\label{H}
\hat{H}=\hslash\lambda(\hat{a}^\dag\hat{a}\hat\sigma_z+\hat\sigma_+\hat\sigma_-),
\end{equation}
where $\hat{a}^\dag$ ($\hat{a}$) is the creation (annihilation) operator of photons, $\hat\sigma_+=\ketbra{e}{g}$ and $\hat\sigma_-=\ketbra{g}{e}$
($\hat\sigma_z=\hat\sigma_+\hat\sigma_--\hat\sigma_-\hat\sigma_+$) are atomic operators, and $\lambda=g^2/\Delta$ is the effective atom-cavity coupling constant. As mentioned before, the interaction Hamiltonian (\ref{H}) does not promote energy exchange between the subsystems. Instead, after an interaction time $t$, a classical phase is added in the coherent field amplitude. Such a phase shift takes opposite values depending on the state of the atom. This process can be summarized as \cite{NLMSHaroche}
\begin{equation}\label{int}
\ket{g}\ket\alpha\rightarrow\ket{g}\ket{\alpha\e^{i\lambda{}t}},\qquad\ket{e}\ket\alpha\rightarrow\e^{-i\lambda{}t}\ket{e}\ket{\alpha\e^{-i\lambda{}t}}.
\end{equation}

After spending an interaction time $t=\pi/2\lambda$ in each of two cavities the atom $i$ ($i=1,2$) and respective cavities are left in the state
\begin{equation}\label{step1}
\ket{\psi}_i=\shalf(\ket{g}_i\ket{\beta,\beta}-\ket{e}_i\ket{{-}\beta,{-}\beta}),\quad(i=1,2),
\end{equation}
where $\beta=i\alpha$. Hence, the state of the whole system, just before the atoms be let to cross the fifth cavity, is given by
\begin{eqnarray}\label{whole}
\ket{\psi}=\ket{\psi}_1\otimes\ket{\psi}_2&=&\half(\ket{gg}\ket{\beta,\beta,\beta,\beta}-\ket{ge}\ket{\beta,\beta,{-}\beta,{-}\beta}\nonumber\\
&&-\ket{eg}\ket{{-}\beta,{-}\beta,\beta,\beta}+\ket{ee}\ket{{-}\beta,{-}\beta,{-}\beta,{-}\beta}).
\end{eqnarray}

In order to achieve the CTECS generation, we need now to perform a controlled phase gate (CPG) operation acting on the atomic Hadamard basis $\ket{\pm}=\shalf(\ket{g}\pm\ket{e})$, followed by atomic state measurement. The desired CPG is represented by the unitary operator $\hat{U}_{\rm CPG}$ whose action in the Hadamard basis is
\begin{eqnarray}\label{UCPG}
&&\hat{U}_{\rm CPG}\ket{{++}}=-\ket{{++}},\nonumber\\
&&\hat{U}_{\rm CPG}\ket{{+-}}=\ket{{+-}},\nonumber\\
&&\hat{U}_{\rm CPG}\ket{{-+}}=\ket{{-+}},\nonumber\\
&&\hat{U}_{\rm CPG}\ket{{--}}=\ket{{--}}.
\end{eqnarray}
The transformation (\ref{UCPG}) may be implemented by using a fifth cavity (see Fig. {\ref{setup}}) as we are now going to explain. The atoms will now simultaneously interact with both the quantum field in the fifth cavity and an externally applied classical field. The respective Hamiltonian in the rotating wave approximation is \cite{Solano}
\begin{equation}\label{Hext}
\hat{H}_{\rm f}=\hslash\omega_0\hat{a}^\dag\hat{a}+\half\hslash\omega\hat\Sigma_z+\hslash{}g'(\hat{a}\hat\Sigma_++\hat{a}^\dag\hat\Sigma_-)+\hslash\Omega(\hat\Sigma_+\e^{-i\omega_ct}+\hat\Sigma_-\e^{i\omega_ct}),
\end{equation}
where $\hat\Sigma_z=\hat\sigma^z_1+\hat\sigma^z_2$, $\hat\Sigma_\pm=\hat\sigma^\pm_1+\hat\sigma^\pm_2$, $\delta=\omega-\omega_0$ is the detuning between the frequencies of the cavity field and atoms, $g'$ is the atom-cavity coupling constant and the atoms are set to be resonant with the classical field ($\omega=\omega_c$). In the strong driving regime $\Omega\gg\{g',\delta\}$, the interaction Hamiltonian is written as \cite{Zhengclas1,Zhengclas2,XiuC}
\begin{equation}\label{HI}
\hat{H}^I_{\rm f}=\half\hslash{}g'(\hat{a}\e^{i\delta{}t}+\hat{a}^\dag\e^{-i\delta{}t})\hat\Sigma_x,
\end{equation}
where $\hat\Sigma_x=\Sigma_{+}+\Sigma_{-}$. Setting $\delta{}t=2\pi$, it can be shown that the strong classical field causes the photon number dependent Stark-shifts to be canceled, letting the fifth cavity insensitive to both cavity decay and thermal field, so that the resulting effective Hamiltonian reads \cite{Zhengclas1,Zhengclas2}
\begin{equation}\label{Heff}
\hat{H}^{\rm eff}_{\rm f}=\hslash\Omega\hat\Sigma_x+\half\hslash\chi\hat\Sigma^2_x,
\end{equation}
where $\chi={g'}^2/2\delta$. This Hamiltonian is suitable for implementation of the {\rm CPG} (\ref{UCPG}) as previously demonstrated in \cite{Zhengclas1,Zhengclas2,ZouCPG}. Following their proposal, it can be shown that in our system, if the atoms spend a time of flight in the fifth cavity fulfilling $t_{\rm f}=\pi/\chi{}$ and $t_{\rm f}=(2k+1/2)\pi/\Omega{}$, for integer $k\geq 1$, the following transformation is performed\newpage
\begin{eqnarray}\label{UCPGHeff}
&&\hat{U}_{\rm CPG}\ket{gg}=\half(\ket{gg}-\ket{ge}-\ket{eg}-\ket{ee}),\nonumber\\
&&\hat{U}_{\rm CPG}\ket{ge}=\half(-\ket{gg}+\ket{ge}-\ket{eg}-\ket{ee}),\nonumber\\
&&\hat{U}_{\rm CPG}\ket{eg}=\half(-\ket{gg}-\ket{ge}+\ket{eg}-\ket{ee}),\nonumber\\
&&\hat{U}_{\rm CPG}\ket{ee}=\half(-\ket{gg}-\ket{ge}-\ket{eg}+\ket{ee}),
\end{eqnarray}
where $\hat{U}_{\rm CPG}=\exp({-iH^{\rm eff}_{\rm f}t_{\rm f}/\hslash})$, with $H^{\rm eff}_{\rm f}$ being the effective Hamiltonian (\ref{Heff}). The transformation (\ref{UCPGHeff}) causes the global state of the system (\ref{whole}) to evolve to
\begin{eqnarray}
\ket{\psi'}=\half(\ket{gg}\ket{{\rm CLUSTER}^+_\beta}-\ket{ge}\ket{{\rm C}^+_\beta}-\ket{eg}\ket{{\rm C}^-_\beta}+\ket{ee}\ket{{\rm CLUSTER}^-_\beta}).\nonumber\\
\end{eqnarray}

The atoms are then detected in the state-selective field ionization counters $D_1$ and $D_2$ (see Fig. \ref{setup}). It is worthwhile to notice that, irrespective to the outcome of the atomic measurement, one \emph{always} ends up with a spatially-separated four-cavity cluster-type entangled coherent state belonging to $\mathcal{B}^\alpha_{\rm C}$. For instance, in the events where atoms are detected in the state $\ket{gg}$, one ends up with the state
\begin{eqnarray}\label{cluster}
\ket{{\rm CLUSTER}^+_{\beta}}&=&\half(\ket{\beta,\beta,\beta,\beta}+\ket{\beta,\beta,{-}\beta,{-}\beta}+\ket{{-}\beta,{-}\beta,\beta,\beta}\nonumber\\
&&-\ket{{-}\beta,{-}\beta,{-}\beta,{-}\beta}).
\end{eqnarray}
Moreover, one may check that other initial preparations like $\ket\psi_{\rm af}=\linebreak\ket{gg}\ket{\alpha,\alpha,\alpha,{-}\alpha}$, $\ket\psi_{\rm af}=\ket{gg}\ket{\alpha,{-}\alpha,\alpha,\alpha}$, and $\ket\psi_{\rm af}=\ket{gg}\ket{\alpha,{-}\alpha,\alpha,{-}\alpha}$ suffice to generate the other elements in the basis $\mathcal{B}^\alpha_{\rm C}$. Conversely, this may also be achieved after the generation by letting atoms to cross each cavity, thus adding classical phases in the coherent field amplitude, which constitute bit-flip operations.

Cluster states are well known for their applications in measurement-based quantum computing, i.e, one way quantum computing \cite{one-way1,one-way2,one-way3}. In the case of qubits, an initial cluster state is prepared and a set of carefully chosen measurements on each qubit is then performed in order to achieve the desired computation. As we would like our cavity QED implementation of the CTECS to be useful in similar measurement-based schemes, we need to consider a way of measuring each cavity field in the basis $\{|\alpha\rangle,|{-}\alpha\rangle\}$. Such measurement scheme is due to J. Lee et al \cite{reciprocation}, and we present it here for completeness. The scheme relies on the choice of measurement operators $\hat{P}_{\alpha}=|\alpha\rangle\langle\alpha|$ and $\hat{Q}_{\alpha}=\id-\hat{P}_{\alpha}$. One can easily check that these operators satisfy the completeness equation $\hat{P}_{\alpha}^\dag\hat{P}_{\alpha}+\hat{Q}_{\alpha}^\dag\hat{Q}_{\alpha}=\id$ which express the fact that the probabilities associated with each outcome sum to one \cite{NielsenChuang}. The way to project the cavity field to the state $|\alpha\rangle$ is to apply the displacement operator $\hat{D}({-}\alpha)$ realizing the transformation $\hat{P}_{\alpha}\rightarrow|0\rangle\langle 0|$, followed by a deliberately lowering of the cavity quality factor allowing the field to leak out the cavity. In the events in which no photons are detected outside the cavity, the field state is projected to $|\alpha\rangle\langle\alpha|$ effectively realizing the measurement $\hat{P}_{\alpha}$. In order to project the field state to $|{-}\alpha\rangle\langle{-}\alpha|$, one must follow the same procedure just changing the sign of $\alpha$. Unfortunately, this scheme is intrinsically probabilistic what seems to put a limit in how well one can measure the cavity field in the basis $\{|\alpha\rangle,|{-}\alpha\rangle\}$. We think that new schemes are to be proposed in the future since coherent states have been extensively considered in the fields of quantum optics, and more recently quantum computing.

It is worth mentioning that this scheme is not limited to quantum optical settings. Actually, Jaynes-Cummings models and its different generalizations have been shown to be implemented in many interesting condensed matter systems such as superconducting junctions \cite{Blais}. Our ideas may in principle be well suited for implementation in such systems.

We now present a few considerations about the feasibility of the scheme in an actual system consisting of Rydberg atoms and microwave cavities \cite{harocherev1,harocherev2}. First, we should notice that the preparation of coherent states in microwave cavities is basically the injection of a classical driving field prior to the passage of the atom. This may be performed with ease nowadays \cite{harocherev1,harocherev2}. Regarding the atoms, Ramsey zones are known to perform the initial rotations also with good fidelity in the current status of cavity QED experiments \cite{harocherev1,harocherev2}. Thus, the initial preparation of the state of the system is not too demanding. Now, the most important aspect to be investigated is the comparison between the total time the atoms spend to cross all cavities with the relaxation times involved in the problem. Each cavity has a photon storage time $T_r=130$ ms \cite{harochelong1,harochelong2}, and the radiative times $T_{at}$ for Rydberg atoms are of the order of $30$ ms \cite{harocherev1,harocherev2}. Having neglected the spatial separation between the cavities, the generation protocol takes a time $T=2\times\pi/2\lambda+\pi/\chi$, that for $g=g'=2\pi\times25$ kHz, and $\Delta=2\pi\times8g$, gives $T\approx1.045$ ms. We can see that $T_r/T\approx124$ and $T_{at}/T\approx29$ what indicates that the implementation of our scheme may be feasible in the present state-of-art in cavity QED setups. However, we would like to emphasize that a more complete analysis would demand the study of time evolution of the initial state of the system (five cavities plus two atoms) under the presence of dephasing and relaxation mechanisms. Also, it is not in the scope of the present paper to study decoherence of the CTECS after completion of the generation protocol. These are interesting problems which deserve a complete study afterwards. The main goal of the present paper is to propose the CTECS and to present a generation protocol using well-established cavity QED setups.

As a final remark, when thinking of potential applications in quantum computing, it would be interesting to extend our generation scheme for an arbitrary number of cavities. For this purpose, instead of crossing just two cavities, each of the two atoms should now cross $p$-cavities in line before interacting via Hamiltonian (\ref{Heff}) in the last cavity. The atom-field interaction in each of the $p$-cavities is still dispersive as described by Hamiltonian (\ref{H}). The system initial state should now read $\ket\psi_{\rm af}=\ket{gg}\ket\alpha^{\otimes{}p}\ket\alpha^{\otimes{}p}$, and the time of flight in each of the $p$-cavities is still given by $2\times\pi/2\lambda$.  Before crossing the last cavity, the state of the atom $i$, and the respective $p$-cavities would be
\begin{equation}\label{stepp}
\ket{\psi}_i=\shalf(\ket{g}_i\ket\beta^{\otimes{p}}+(-i)^p\ket{e}_i\ket{{-}\beta}^{\otimes{p}}).
\end{equation}

After an interaction time $\pi/\chi$ in the last cavity, and subsequent measurement of the atoms, for example, in $\ket{gg}$,  the state of the $2p-$cavities collapses to
\begin{eqnarray}
\ket{{\rm CLUSTER}^+_{\beta,p}}&=&\frac{1}{\sqrt{N_p^+}}(\ket\beta^{\otimes{p}}\ket\beta^{\otimes{p}}-(-i)^p\ket\beta^{\otimes{p}}\ket{{-}\beta}^{\otimes{p}}-(-i)^p\ket{{-}\beta}^{\otimes{p}}\ket\beta^{\otimes{p}}\nonumber\\
&&-(-1)^p\ket{{-}\beta}^{\otimes{p}}\ket{{-}\beta}^{\otimes{p}}),
\end{eqnarray}
where $N_p^+=2\{2+[1-(-1)^p]e^{-4p|\alpha|^2}\}$. This is an example of CTECS involving $2p$ elements with $p$ an arbitrary positive integer.

In summary, we have proposed a new type of entangled multipartite states involving coherent states (CTECS) which generalize the usual cluster states. Such states are seen to form an interesting basis for expanding general $4$-qubit states since all elements in the basis possess the same amount of entanglement. We believe that such a basis may find applications in future quantum teleportation protocols based on the just proposed CTECS. We proposed an actual implementation in a quantum optical setup consisting of Rydberg atoms and microwave cavities, and we found that our ideas might, in principle, be implemented within current technology.

\section*{Acknowledgements}

This work is supported by CNPq (Conselho Nacional de
Desenvolvimento Cient\'ifico e Tecnol\'ogico) and FAPESP (Funda\c c\~ao de Amparo \`a Pesquisa do
Estado de S\~ao Paulo), Brazil. This work also is linked with CEPOF (Centro de
Pesquisa em \'Optica e Fot\^onica , FAPESP).

\newpage

%%%%%%%%%%%%%%%%%%%%%%%%%%%%%%%%%%%%%%%%%%%%%%%%%%%%%%%%%%%%%%%%%%%%%%%%%%%%%%%%%%%%%%%%%%%%%%%%%%%%%%%%%%%%%%%%%%%%%%%

\newpage

% \appendix

% \section{}
% \label{}

%\begin{thebibliography}{00}


% notes:

% \bibitem{label2}
% If there is a note, it should come last:
% \bibitem{label3} \note
% \end{subbibitems}

%\bibitem{}

%\end{thebibliography}

\end{document}